\documentclass[prd,showpacs,amsmath,showkeys,twocolumn,floatfix,amssymb, preprintnumbers, nofootinbib, superscriptaddress]{revtex4}

\usepackage{graphicx}
\usepackage{color}

\newcommand{\be}{\begin{equation}}
\newcommand{\ee}{\end{equation}}

\newcommand{\nn}{\nonumber}

\newcommand{\beq}{\begin{equation}}
\newcommand{\eeq}{\end{equation}}
\newcommand{\beqa}{\begin{eqnarray}}
\newcommand{\eeqa}{\end{eqnarray}}
\newcommand{\bd}[1]{ \mbox{\boldmath $#1$}}

\usepackage{bm}  
\renewcommand*{\vec}[1]{\ensuremath{\bm{\mathrm{#1}}}}

\begin{document}

\title{Charmonium meson and hybrid radiative transitions}

\author{Peng~Guo}
\email{pguo@jlab.org}
\affiliation{Thomas Jefferson National Accelerator Facility,  
Newport News, VA 23606, USA}
\affiliation{Center For Exploration  of Energy and Matter, Indiana University, Bloomington, IN 47408, USA.}

\author{Tochtli Y\'epez-Mart\'inez}
\affiliation{Center For Exploration  of Energy and Matter, Indiana University, Bloomington, IN 47408, USA.}

\author{Adam~P.~Szczepaniak}
\affiliation{Thomas Jefferson National Accelerator Facility,  
Newport News, VA 23606, USA}
\affiliation{Center For Exploration  of Energy and Matter, Indiana University, Bloomington, IN 47408, USA.}
\affiliation{Physics Department, Indiana University, Bloomington, IN 47405, USA}

\preprint{JLAB-THY-14-1851}

\date{\today}

\begin{abstract}
We consider the non-relativistic limit of the QCD Hamiltonian in the Coulomb gauge, to describe radiative
transitions between  conventional charmonium states and from the
lowest multiplet of $c\bar{c}$ hybrids to charmonium mesons. The results
are compared to potential quark models and lattices calculations. 
\end{abstract}


\maketitle

\section{Introduction}
\label{intro} 

It has long been stipulated that excitation of the gluon field would appear in the spectrum of hadrons.
 Hybrid resonances, {\it i.e.} states that contain both quark and gluon excitations, were considered in various models,   \cite{Horn, Isgur, Simonov, Adam-Eric-1996, Buisseret, Brau}, 
 and recent lattice simulations \cite{glueball-1, glueball-2, glueball-3} have provided solid theoretical evidence for such states. 
   Moreover, in recent years several new states, in particular in the charmonium spectrum have been discovered  possibly 
      including  a hybrid resonance, the $Y(4260)$. 
   Conventional heavy quarkonia are well described by non-relativistic
 QCD \cite{Cornell}. Thus it is reasonable to expect that hybrids containing heavy quarks could be treated in a similar way, {\it i.e.} by  considering gluon excitations in presence of slowly  moving
 quarks.  In physical gauges, {\it e.g.} the Coulomb gauge, dynamical
 gluons can be separated from the instantaneous Coulomb-type forces
 that act between color charges \cite{Lee, Llanes, General,
   Peng-Adam-1-2008, Peng-Adam-2-2008, Feuchter1, Feuchter2}. 
 The non-abelian Coulomb potential is expected to be responsible for
 binding and confinement \cite{Zwanziger, Greensite} while the remaining, transverse gluon excitations could  contribute to the spectrum. 
   
To a good approximation heavy quarks interact with photons as bare Dirac particles. Thus 
 radiative transitions can be used to  explore quarkonium dynamics.   We assume that this phenomenology can be extended to  quarkonium hybrids.  Over the years several radiative transitions involving charmonia have been
measured ~ \cite{glueball-rev1, glueball-rev2, Eric-2006} and 
 extensive theoretical studies were performed 
\cite{Eric-2005, Eichten-1978, Eichten-1980, Eichten-2002}. More recently lattice gauge simulations have become available 
 \cite{Dudek-2006, Dudek-2008}  and these also  include predictions  for transitions involving hybrid mesons
 \cite{Dudek-2009, Dudek-2011}. 

 In this work we focus on radiative transitions
involving lowest mass conventional charmonia and the lowest mass multiplet of charmonium hybrids. 
The ordinary $c\bar c$ states we consider have  quark orbital angular momentum and spin restricted to the lowest values, of $L,S=0,1$ that  
  results in states with angular momentum, parity and charge
 conjugation, $J^{PC} = 0^{-+}, 1^{--}, 1^{+-}, (0,1,2)^{++} $. 
 In the non-relativistic, Coulomb gauge QCD the lowest mass charmonium hybrid multiplet is predicted to 
 contain a color-octet   $c\bar c$ pair with  $J_q^{P_qC_q} = 0^{-+}$ or $1^{--}$ corresponding 
  to the total quark-antiquark spin  $S=0$ and $S=1$, respectively, coupled to  a single quasi-gluon. 
  This physical, transverse gluon is predicted to have quantum numbers, $J_g^{P_g C_g} = 1^{+-}$. The unusual, positive parity
   of the gluon originates from the non-abelian nature of the Coulomb
   interactions \cite{Peng-Adam-1-2008, Peng-Adam-2-2008}. 
 Coupling of the $c\bar c$ and the gluon produces a multiplet containing four hybrid states, with overall quantum numbers of $J^{PC} = 1^{--}, (0,1, 2)^{-+}$. 
 This four state multiplet has been recently identified in lattice simulations, both in the heavy and light quark sectors. 
 It includes the exotic state with $J^{PC}=1^{-+}$ and three states with non-exotic quantum numbers,  
$1^{--}, 0^{-+}, 2^{-+}$. The gluon content of the former was identified trough determination of   matrix elements of operators containing gluon fields  \cite{Dudek-2006,
  Dudek-2009, Dudek-2011}

The paper is organized as follows.  In Section~\ref{Ham} we detail the Coulomb gauge approach to 
conventional charmonium radiative transitions and to transitions involving hybrid mesons. 
We discuss the basis states for ordinary $c\bar c$ mesons and
$c\bar c g$ hybrids and the corresponding 
  transition matrix elements. 
  In Section ~\ref{NR:CM} 
a multipole analysis of the radiative transitions is presented. We also discuss current matrix elements involving states of identical 
 charge conjugation. These vanish when photon couples to both the quark and the antiquark but are in general finite when the current operator acts on a single quark. They are well defined within the model and have also been computed on the lattice. 
  Summary and outlook are given in Section~ \ref{SO} and  all details of derivations are given in the appendices.

\section{\label{Ham} Quarkonium states in the Coulomb gauge}

The QCD Hamiltonian $H_{QCD}$, which describes 
non-relativistic quarks interacting with (relativistic) gluons can be constructed from the full QCD
Hamiltonian in the Coulomb gauge by applying Foldy-Wouthuysen
transformation \cite{Foldy-1978}.  This Hamiltonian was used to study 
 the gluelump spectrum \cite{Peng-Adam-1-2008} and the
low mass charmonia and bottomonia including hybrids 
\cite{Peng-Adam-2-2008}. In addition to the strong interaction part, here we also consider the minimal coupling of the photon to the quarks, which in the non-relativistic limit 
is given by 
\begin{equation} 
H_{QED}  = \frac{e_q}{2m} \int d {\bf x} \Psi^{\dagger}
({\bf x}) \beta [2i {\bf A}_{\gamma} ({\bf x}) \cdot \nabla 
-{\bf \Sigma } \cdot
{\bf B}_\gamma({\bf x })] \Psi({\bf x})  \label{qed} 
\end{equation} 
where ${\bf A_\gamma}$ and ${\bf B}_\gamma$ are the photon vector potential and magnetic field, respectively. The quark fields are related to particle operators by 
\begin{equation} 
\Psi^i  ({\bf x}) = \sum_{\lambda=\pm1/2 } \int \frac{ d {\bf k} }{ (2\pi)^3 }
e^{i {\bf k} \cdot {\bf x} } 
[u_\lambda b({\bf k},\lambda,i)  + v_\lambda d^{\dagger}(-{\bf k},\lambda,i)] 
\end{equation} 
 with $u,v$ being the Dirac spinors in the non-relativistic limit. Given an (approximate) solution of the Schr\"odingier equation 
  \begin{equation} 
 H_{QCD} |N[c\bar c]\rangle = E_N |N [c\bar c]\rangle
\end{equation} 
within the Fock sector containing only the heavy quark-antiquark pair  the QED interaction of Eq.~(\ref{qed}) determines 
 the radiative transition matrix element, 
\begin{equation} 
{\cal M}_{N \to N'\gamma}  \propto \langle N' [c\bar c],\gamma| H_{QED} |N [c\bar c]\rangle 
\end{equation} 
between ordinary   charmonia.  In the case of transitions involving hybrids, which are given by solutions of 
\begin{equation} 
  H_{QCD} |N[c\bar c g]\rangle = E_N |N [c\bar c g]\rangle \label{sh} 
 \end{equation}
 in the sector containing in addition to the $c\bar c$ pair a
 transverse quasi-gluon, the radiative transition to an ordinary meson state has to be accompanied by gluon absorption. To lowest order in the heavy quark mass expansion the later is determined by the instantaneous Coulomb interaction that 
   changes the gluon number,  $ \langle c\bar c |H_C| c\bar c g\rangle $. Here  $H_C$ given by 
 
 \begin{equation}{\label{Ham-QCD-QED}}
H_C  = -\frac{g^2}{2} \int d{\bf x} d{\bf  y}
\rho^a ({\bf x}) 
K_{a,b} ({\bf x }, {\bf y }, {\bf A }_g) 
\rho^b ({\bf y}) 
\end{equation} 
 and  $\rho^a ({\bf x}) = \Psi^{\dagger}({\bf x})T^a \Psi({\bf x}) $ is the quark color charge density  and the gluon field ${\bf A}_g$ 
is related to the quasi-gluon particle operators by 
\begin{align} 
{\bf A}^a_g({\bf x}) = 
\int \frac{ d {\bf k}  }{ (2\pi)^3 }
\frac{  e^{i {\bf k} \cdot {\bf x} }    }{\sqrt{2 \omega(k) }} &
 [ 
\vec \epsilon ({\bf k},\lambda)a({\bf k},\lambda,a) \nn\\
   & \!\! +  \vec \epsilon^{\dagger} (-{\bf k},\lambda)a^{\dagger}(-{\bf k},\lambda,a)
 ] ,   \nn\\
\label{ag} 
\end{align}
with  $\lambda$, $a$ being  the
helicity and color indices, respectively and
$\vec \epsilon({\bf k}, \lambda)$  the helicity vectors. 
The quasi-gluon orbitals and the quasi-gluons dispersion function  $\omega_k = \omega(k)$ have been studied elsewhere 
 using a variational model for the QCD vacuum ~\cite{Peng-Adam-2-2008}. In the variational model 
 the Coulomb kernel is replaced by its vacuum expectation value and 
    the operator which changes the gluon number by one becomes, 
\begin{equation}{\label{QCD-pot}}
K_{a,b}  = f^{abc} 
  \int \frac{ d{\bf k} } {(2\pi)^3} \frac{ d{\bf q}  } {(2\pi)^3}
  e^{i {\bf k}{\bf x} -i{\bf q} \cdot {\bf y} }  
{\bf k}  \cdot {\bf A}_g^{c}( {\bf k}-{\bf q}) K^1 (k,q) 
\end{equation}
with the scalar  function $K^1(k,q)$ obtained from a solution of a series of
Dyson-Schwinger equations
~\cite{Adam-Eric-2001, Adam-2004, Hugo-2004, Hugo-2005, Hugo-2006,
  Hugo-2007, Hugo-Adam-2008}.  
  The model
 has been used successfully  \cite{Krupinski 1, Krupinski 2} in the
 study of excited adiabatic potentials between static quarks
 \cite{Morningstar-1998},   which can be used to determine  the single gluon orbitals
 in Eq.~(\ref{ag}).   Combining Eqs.~(\ref{qed},\ref{Ham-QCD-QED})
 leads to an effective operators for radiative transitions between hybrid and ordinary quarkonia
\begin{equation} \label{M-hy-mes}
{\cal M}_{N \to N'\gamma}  \propto \langle N' [c\bar c],\gamma| H^{eff}_{QED} |N [c\bar c g]\rangle 
\end{equation} 
where 
\beqa\label{HeffQED}
&&H_{QED}  ^{eff}= \frac{1}{2}  \frac{ H_{C}  H_{QED} 
}{ \Delta E } 
\eeqa
with $1/\Delta E $ representing the Green's function of the $c \bar c$ pair.  In the following we calculate the matrix elements $\mathcal{M}$ and the decay widths
for several hybrid states. As discussed previously, we focus on
the hybrid states containing quark and antiquark angular momentum $L=0,1$ and spin 
$S=0,1$. In particular we investigate  
 transitions involving the  hybrid with exotic quantum numbers $\eta_{c1}(1^{-+})$. This state has
been described by lattice calculations \cite{Dudek-2006, Dudek-2009} 
 and is expected to have a mass around $4.3 \mbox{ GeV}$.

\subsection{\label{mbme}Meson basis and matrix elements.} 
We represent the N-th quarkonium state of spin $J$ and its projection
 $M$, with parity $P$ and charge conjugation $C$ and total momentum 
 ${\bf P}$  as 
\begin{align}\label{QQ-state}
& | {\bf P}; JMPCN \rangle = \sum_{\alpha,  m_1 , m_2 }\int \frac{ d {\bf q}}{(2\pi)^3}
\Psi^{N,\alpha}_{c\bar{c}}(q)  \nn\\
& \times 
\chi_{m_1 , m_2}^{JMPC} (\hat{ {\bf P} }, 
\hat{ {\bf q} }, \alpha)  b^{\dagger} ( {\bf p}_c , m_1, i_1 )\frac{ \delta_{i_1 , i_2} }{\sqrt{N_c}}
d^{\dagger} ( {\bf p}_{\bar{c}} , m_2, i_2 )  |0\rangle . \nn\\
\end{align}
Here $\alpha = (L,S)$, and $q$ is the magnitude of relative momentum between quark and antiquark. ${\bf p}_c = \frac{{\bf P}}{2} + {\bf q}$ and ${\bf p}_{\bar{c}} = \frac{{\bf P}}{2} - {\bf q}$  are the quark and antiquark momenta respectively. The meson spin-orbital wave function is written using the $L-S$ coupling scheme  with $L$, and $S$ the orbital angular momentum and spin of the quark-antiquark, respectively, 
\begin{align}
&\chi_{m_1 , m_2}^{JMPC} (\hat{ {\bf P} }, 
\hat{ {\bf q} }, \alpha) =
\sum_{M_S , M_L} Y_{LM_L}(\hat{ {\bf q} })  \langle \frac{1}{2} m_1 ; \frac{1}{2} m_2| S M_S \rangle \nonumber \\
&\times \langle S M_S ; L M_L | JM \rangle
\frac{1+C(-1)^{L+S} }{2}  \frac{1+P(-1)^{L+1}}{2}  .
\end{align}
The states are normalized according to 
\begin{align}\label{statenorm}
&\langle {\bf P}^{\prime}; 
J ^{\prime}M ^{\prime}P ^{\prime}C^{\prime} N ^{\prime} 
| {\bf P}; JMPCN \rangle\nn\\
& \quad  \quad = 2E_{c\bar{c}}
(2\pi)^3 \delta^{3}({\bf P}- {\bf P}^{\prime}) \delta_{J J^{\prime}} 
\delta_{M M^{\prime}} \delta_{P P^{\prime}} \delta_{C C^{\prime}} 
\delta_{N N^{\prime}} .
\end{align}

As mentioned before the meson-to-meson radiative transitions are calculated
 with the minimal coupling of the photon to the quarks, {\it cf. }
Eq.~(\ref{qed}).  Explicitly, the matrix elements are given by
\begin{align}{\label{M to M}}
& \mathcal{M}_{N \to N' \gamma} \nonumber \\
&= -\frac{e_q }{2m_q (2\pi)^3}
\int  d {\bf q}  d {\bf q}^{\prime} 
\Psi_{c\bar{c}}^{N,\alpha}(q)
\Psi_{c\bar{c}}^{N^{\prime},\alpha^{\prime}} (q^{\prime}) \nonumber \\
&\times \sum_{m_1, m_2,m_1   ^{\prime} , m_2 ^{\prime}}  
\chi_{m_1 , m_2 } ^{* JMPC} (\hat{ {\bf  q}  },\alpha)
\chi_{m^{\prime}_1 , m^{\prime}_2 }^{J^{\prime}M^{\prime}P^{\prime}C^{\prime}}
 (\hat{ {\bf q}^{\prime}  },\alpha^{\prime}) \epsilon ({\bf k}_{\gamma}, \sigma_{\gamma})    \nn\\
&\cdot \left [
   \delta({\bf q}^{\prime}- {\bf q}  +\frac{{\bf  k}_{\gamma}}{2})
  \left ( 2 
{\bf q}^{\prime}   
+ i \sigma 
\times {\bf k}_{\gamma}  
\right )_{m_1 , m_1 ^{\prime} } \delta_{m_2 m_2 ^{\prime}}\right.\nn\\
&\ + \left.
  \delta( {\bf q} -{\bf q}^{\prime} +\frac{{\bf k}_{\gamma}}{2})   
\left ( 2 
{\bf q}^{\prime}    + i ( \sigma_{2} \sigma  \sigma_{2})
\times {\bf k}_{\gamma}  
\right )_{m_{2}^{\prime}, m_{2}} \delta_{m_1 m_1 ^{\prime}}\right ]. \nn\\
\end{align}
 
The spin-orbital wave function $\chi^{JMPC}_{m_1 ,
  m_2}(\hat{ {\bf q} }, \alpha)$ for charmonium mesons
$J^{PC}=0^{-+},1^{--}, 1^{+-}, (0,1,2)^{++}$ are
tabulated  in Appendix \ref{A:mhsowf}.

\subsection{{\label{hbme}}Hybrid basis and transition matrix elements}
It is reasonable  to assume that wave function of  hybrids with non relativistic quarks are similar to those of  gluelumps
 which contain static quarks.   In construction of  hybrid wave functions we thus follow the coupling scheme optimized for 
 gluelump studies \cite{Peng-Adam-1-2008}. 
 The $Q\bar{Q}g$ state is obtained by initially coupling the $Q\bar{Q}$ relative angular
momentum $L$ to the total gluon spin $J_g$.  The
resulting angular momentum $j$ is then coupled to the total
quark-antiquark spin $S$ to give the total spin of the hybrid state
$J$. The hybrid state with total spin and it projection, 
 $J$,$M$, parity $P$, charge conjugation
$C$ is then given by
\begin{align}{\label{hwf}}
& |JMPCN\rangle = \sum_{\alpha = (J_g S, L, j)}   
\int \frac{ d  {\bf k} }{(2\pi)^3}  \frac{d  {\bf q} }{(2\pi)^3}
\Psi^{ N,\alpha}_{c\bar{c}g} (k,q)\nn\\
&\times  \sum_{m_1 , m_2 , \sigma}
  \frac{1}{\sqrt{C_F N_c}}
\chi_{m_1 , m_2 , \sigma} ^{JMPC} (\hat{ {\bf k}  },\hat{ {\bf q}  },
\alpha)  \nn\\
&\times
b^{\dagger} ( \frac{{\bf k}}{2}+{\bf q}, m_1 , i_1) T^{a}_{i_1 , i_2} 
d^{\dagger} ( \frac{{\bf k}}{2}- {\bf q}, m_2 , i_2) a^{\dagger} ( - {\bf k}, \sigma , a) |0\rangle. \nn\\
\end{align}
Here ${\bf q}$ is the relative momentum between the quark-antiquark
and ${\bf k}$ is the momentum of the gluon in the overall center of mass frame. The spin-orbital wave function $\chi_{m_1 , m_2 , \sigma} ^{JMPC}
(\hat{ {\bf k}  },\hat{ {\bf q}  }, \alpha)$ describes the
$(\bd{L}+\bd{J}_g)+\bd{S}$ coupling and $\sigma=\pm 1$ represents the
gluon helicity
\begin{align}
&\chi_{m_1 , m_2 , \sigma} ^{JMPC} (\hat{ {\bf k}  },\hat{ {\bf q}  },\alpha) 
= \sqrt{\frac{2J_g + 1}{4\pi}}  
\frac{1+C(-1)^{L+S+1}}{2} \nn\\
&\times 
 \sum_{M_S , M_L , M_{g}, m}   Y_{LM_L}({\bf q})
\langle \frac{1}{2} m_1 ,  \frac{1}{2} m_2  | S M_{S}  \rangle \nn\\
&\times 
\langle J_g M_g ,  L M_L  | j m  \rangle  \langle j m ,  S M_S  | J M \rangle
\nn\\
&\times
 \frac{(-1)^{J_g} }{\sqrt{2}}D^{* J_g}_{M_g , -\sigma}(\hat{  {\bf k} }) \left [\delta_{\sigma, 1} +
P(-1)^{J_g +L+1} \delta_{\sigma, -1} \right ] .
\nn\\
\end{align}
The parity and charge conjugation are given by 
\begin{equation}
P =\xi (-1)^{J_g +L+1} ~, 
C = (-1)^{L+S+1} ~,
\end{equation}
respectively.  Here $\xi =+1$ corresponds to the TM (natural parity)
and $\xi =-1$ for TE (unnatural parity) gluon state that are given be 
   $| \sigma=+1 \rangle + \xi | \sigma=-1 \rangle$ combinations of
gluon helicity states. 
As expected, both P and C are a product of the 
$Q\bar{Q}$  and gluon parity and charge conjugation and are given by
\begin{align}
P_q &= (-1)^{L+1} ~, P_g = \xi (-1)^{J_g}\nn\\
C_q &= (-1)^{L+S} ~, C_g = -1.
\end{align}
The state is normalized in the same way as the normalization of
conventional meson state in Eq.(\ref{statenorm}).  For the lowest four hybrids we are considering \cite{Peng-Adam-2-2008},
  $(L, J^{P_g C_g}_g) =(0,1^{+-})$, which correspond to the gluon
in the TE mode. Coupling the TE gluon with the color octet $Q\bar{Q}$ state
in $L=0$, produces a hybrid state with the
intermediate angular momentum ${\bf j} = {\bf L} + {\bf J}_g=1$. 
Adding the quark spin $S=0,1$, and ignoring hyperfine splitting 
  we obtain four low lying hybrids with quantum numbers, $J^{PC}=1^{--}$ for 
$S=0$ and $ J^{PC}=0^{-+}, 1^{-+}, 2^{-+}$ for $S =1$.  It is worth 
noting that the hybrid with exotic quantum numbers $1^{-+}$ appears in
this lowest multiplet and is predicted to have the $Q\bar{Q}$ pair in
spin-1.

The matrix elements  for the hybrid-to-meson radiative
transition are given by
\begin{widetext}
\begin{align}{\label{H to M}}
\mathcal{M}_{N \to N' \gamma}
&= \frac{e_q g^2}{4m_q} \frac{N_c \sqrt{C_F}}{\sqrt{2}} 
\int \frac{d {\bf k }}{(2\pi)^3}  \frac{d {\bf q}}{(2\pi)^3} \frac{d {\bf q}^{\prime}}{(2\pi)^3} 
\Psi_{c\bar{c}g}^{N,\alpha}(k,q)
\Psi_{c\bar{c}}^{N^{\prime},\alpha^{\prime}} (q^{\prime})
\frac{1}{\sqrt{  \omega_{k_{\gamma}} \omega_k}\Delta E}\nn\\
&  \times  
\sum_{m_1 m_2 \sigma}\sum_{m_1   ^{\prime} m_2 ^{\prime}}  
\chi_{m_1 , m_2 , \sigma} ^{* JMPC} (\hat{ {\bf k}  },\hat{ {\bf  q}  },\alpha)
\chi_{m^{\prime}_1 , m^{\prime}_2 }^{J^{\prime}M^{\prime}P^{\prime}C^{\prime}}
 (\hat{ {\bf q}^{\prime}  },\alpha^{\prime})
\int  d {\bf q}_{g} 
{\bf q}_g  \cdot \epsilon^{*} ( -{\bf k},\sigma )
K^{(1)}(|\frac{{\bf k}}{2}+{\bf q}_g|, |\frac{{\bf k}}{2} -{\bf q}_g|)  \nn\\
& \times  \epsilon( {\bf k}_{\gamma}, \sigma_{\gamma})    \cdot  \left\{ \delta( {\bf q} +{\bf q}_g  
-{\bf q}^{\prime} -\frac{{\bf k}_{\gamma}}{2}) \left
[2 
({\bf q}^{\prime} + \frac{ {\bf k} }{4} - \frac{ {\bf q}_g }{2})  
+ i \sigma
\times  {\bf k}_{\gamma}  
\right ]_{m_1 , m_1 ^{\prime}}  \delta_{m_2 m_2 ^{\prime}}\right.\nn\\
& \quad \quad \quad \quad  \quad     +\left.  
 \delta( {\bf q} +{\bf q}_g  -{\bf q}^{\prime} 
+\frac{{\bf k}_{\gamma}}{2})  \left 
[2  
({\bf q}^{\prime} - \frac{ {\bf k} }{4} - \frac{ {\bf q}_g }{2})  
 + i  (\sigma_{2} \sigma \sigma_{2})  \times
  {\bf k}_{\gamma} 
 \right ]_{m_{2}^{\prime} , m_{2}} \delta_{m_1 m_1 ^{\prime}}\right\}~. 
\end{align}
\end{widetext}
 The explicit form of the spin-wave functions are summarized  in the
Appendix \ref{A:mhsowf}.  In Fig. \ref{FIG-Heff},  we illustrate one of the four possible ways of coupling the photon 
 to a quark line. 
\begin{figure}[h]
\begin{center}
\includegraphics[width=8.0cm, height=4.5cm]{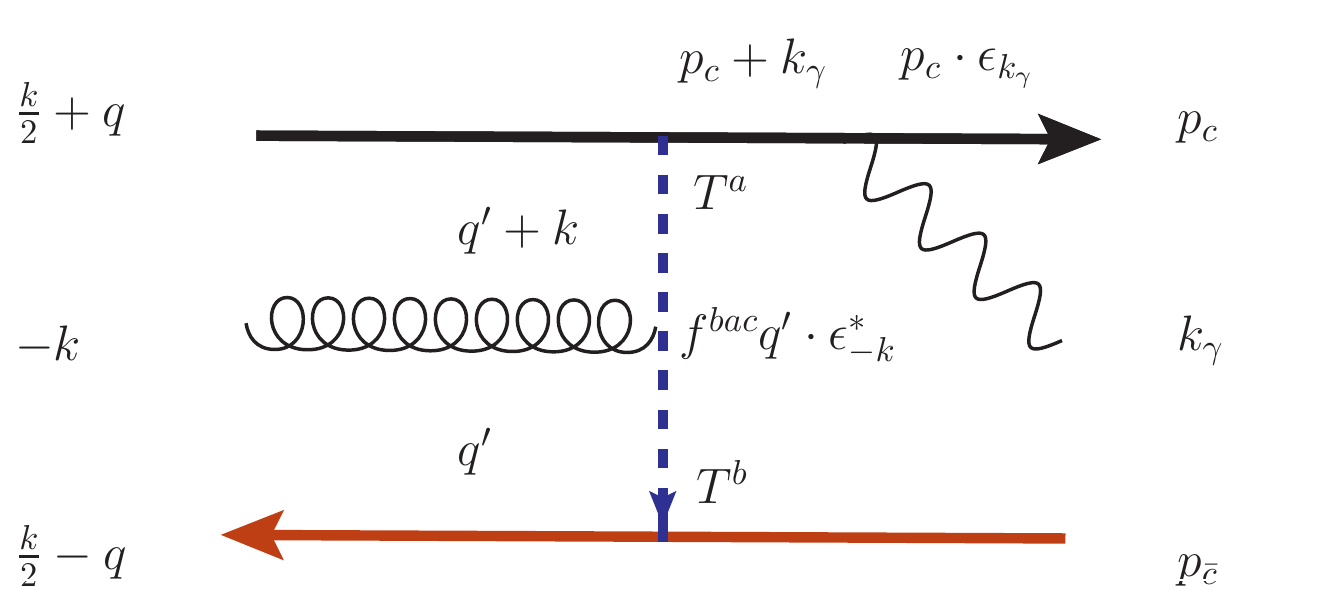}
\end{center}
\caption{Diagrammatic representation of one possible configuration for hybrid to
                  meson  radiative transitions contributing to the matrix element in Eq.~(\ref{H to M}). } \label{FIG-Heff}
\end{figure}


\section{\label{NR:CM}Radiative transitions: Numerical results and discussion}

\subsection{ Conventional mesons}

We have considered a total of fifteen transitions between  conventional charmonia.  Even though some of the
transitions considered here vanish due to charge conjugation, we
investigate the underlying matrix elements with photon attached to only one of the quarks. 
 Some of these C-violating results can be
compared with lattice results reported in \cite{Dudek-2009}, and 
others constitute our predictions.  Using the model described in
Sec.~\ref{Ham}, we present below the final expressions  for
the matrix elements and decay widths computed from 
\begin{align}
&\Gamma(N\to N'\gamma)\nn\\
& \quad =\int d\Omega_{\gamma}\frac{1}{32\pi^{2}}
\frac{{k}_{\gamma}}{m^{2}_{N}}\frac{1}{2J_N
  +1}\sum_{\sigma_{\gamma} , M_N , M_{N'}}
|\mathcal{M}_{N\to N'\gamma} |^{2} .\nn\\
\end{align}
A summary of numerical  results 
 is given in Table~\ref{tab:MM}, including ratios of decay widths relative to 
$\Gamma(\chi_{c_2} \to \gamma J/\psi)$, {\it e.g.} $R_{N\to N'} \equiv \Gamma(N \to N'
  \gamma) / \Gamma(\chi_{c_2} \to \gamma J/\psi)$,  which are compared
  to model calculations from \cite{Eric-2005}. 
We also discuss the transition amplitudes $|\hat{V}|$ and $|\hat{F}_k|$
 introduced in \cite{Dudek-2006, Dudek-2009} in the context of analysis of lattice data. 
 Here  $\hat{F}_k$ represents  either electric, $\hat{E}_k$ or magnetic,  $\hat{M}_k$ multipole  and $\hat{V}$ is 
 the dipole magnetic multipole for the transition involving  a vector and a pseudoscalar meson, 
\begin{align}{\label{FV}}
|\hat{F}|^2&=|\hat{F}_1|^2=\frac{1 }{8 e_q^2 }\sum_{\sigma_{\gamma},M_N, M_{N'}}
 |\mathcal{M}_{N\to N' \gamma}|^2, \nn\\
|\hat{V}|^2&=\frac{  (m_N + m_{N'})^2 }{32 e_q^2 m^2_N k^2_\gamma }\sum_{\sigma_{\gamma},M_N, M_{N'}}
 |\mathcal{M}_{N\to N' \gamma}|^2  .
\end{align}
For the radial wave functions we use a harmonic oscillator approximation 
  with a width parameter $\beta=0.5~\mbox GeV$. This leads to some differences 
 with respect to the other potential-quark results of
  \cite{Eric-2005},
  where a Coulomb plus linear plus and hyperfine interactions were used to compute the wave functions. 
Finally, we calculate the transition amplitudes for charge
conjugation violating  transitions, $|\hat{F}|/2$. The factor
of two is introduced to account for the fact that photon couples to a single quark.
Our findings are summarized below. 
 
\subsubsection{ $   \chi_{c2}(2^{++}) \to h_c (1^{+-})
 \gamma  $ }
 A summary of  recent experimental results on the decays of
charmonium can be found in \cite{Gu-1999}. 
To the best of our knowledge, however,  this transition has not been
measured. 
 It  corresponds to a magnetic dipole, which in general are expected to be weaker 
 than the electric dipole transition. 
The matrix element corresponding to the dominant, 
 $M_1$ transition is  given by 
\begin{equation}
\mathcal{M}_{N\to N'\gamma}  = -\frac{e_q }{m_q}
\frac{3 i}{ 4\pi } 
\epsilon^{* ij}_{M}   
 \left [ {\bf k}_{\gamma} \times \epsilon ( {\bf k}_{\gamma},\sigma_{\gamma}) \right ]^{i}  \epsilon^j_{M^{\prime}}  
\mathcal{A} , 
\end{equation}
where $m_q$ and $e_q = (2/3) \sqrt{4\pi \alpha_{em}}$ are the charm quark mass and charge, respectively.  
Here $\epsilon^{ij}_M$ and $\epsilon^i_M$  are the spin-2 and spin-1
polarization vectors, respectively and the scalar function
$\mathcal{A}$ 
is given in the Appendix \ref{A:mmre}.
Using the harmonic oscillator wave function we obtain 
$\Gamma = 0.1 \mbox{ keV}$. 
The difference with respect to the expressions
given in \cite{Eric-2005} can be traced to an
intrinsic ambiguity in normalization of the wave functions,  
{\it i.e} the difference is  of the order of 
$E^{c \bar c}_f / M_i ^{c \bar c}-1$. 
A more extended theoretical and experimental report on heavy
quarkonium physics is given in \cite{cern-col}, where the effects of higher order relativistic corrections are discussed.

\subsubsection{\label{2++ to 1--}$ \chi_{c2}(2^{++}) \to J/\psi (1^{--})
 \gamma $}
This tensor-to-vector transition has been
studied in potential-quark models \cite{Eichten-2002, Eric-2005} and also on the 
lattice \cite{Dudek-2009}. 
There is experimental evidence for transitions involving radial excitations of the tensor
states $\chi '_{c2} \to J/\psi \gamma$
and $\chi '' _{c2} \to J/\psi \gamma$, but in this paper we focus on the
ground state tensor, $\chi_{c2}$. The corresponding  matrix element is given by 
\begin{equation}
\mathcal{M}_{N\to N'\gamma}
= -\frac{e_q }{m_{q}}
\frac{\sqrt{3}}{ 2\pi }
 \epsilon^{* ij}_{M} \epsilon^{ i}_{M^{\prime}} 
\epsilon^j ( {\bf k}_{\gamma} \sigma_{\gamma})  \mathcal{D} ,
\end{equation}
with $\mathcal{D}$ given
  in the Appendix \ref{A:mhsowf}. The multipole decomposition,
Eq.~(\ref{Dudek-multipole})  yields an 
  electric dipole $E_1$, magnetic quadrupole $M_2$ and 
electric  octopole $E_3$, with $E_1$ being the leading one. 
 The calculated value for the decay width of $\Gamma =363\mbox{ keV}$ in our model 
    agrees  with experimental data \cite{pdg-2012} and lattice calculations \cite{Dudek-2009}. 
The FermiLab-E760 \cite{E760}, BES-collaboration \cite{BES} and CLEO collaboration \cite{CLEO} have all 
reported this transition.   The PDG \cite{pdg-2012} reports a decay 
width $\Gamma=380\mbox{ keV}$. The potential-quark models
 give a width within the range of $\Gamma \approx 289 -
424\mbox{ keV}$.  The electric dipole transition amplitude value
from lattice calculations is $|\hat{F}|=|\hat{E}_1|=1.97~\mbox{GeV}$
and it is obtained by extrapolating the electric dipole form factor to the
physical photon point $\hat{E}_1(Q\to 0)=\hat{E}_1$. All results are
summarized in Table \ref{tab:MM}.

\subsubsection{\label{1+- to 1++}$ h_{c}(1^{+-}) \to \chi_{c1} (1^{++})
 \gamma $}
To the best of our knowledge there is no experimental information about this transition. 
The only observed transition between the $ h_{c}(1^{+-}) $ and another
$c\bar c$ meson is $h_{c}(1^{+-}) \to \eta_{c} (0^{-+}) \gamma$
\cite{pdg-2012}, which we discuss later. 
 The matrix element for this transition is given by 
\begin{equation}
\mathcal{M}_{N\to N'\gamma}
=
\frac{e_q }{m_q}
\frac{3 }{ 4\pi \sqrt{2} } \varepsilon_{ijk} 
\varepsilon_{ilm}
\epsilon^{* j}_{M} 
\epsilon^k_{M^{\prime}} 
{\bf k}^l_{\gamma}
\epsilon^m ( {\bf k}_{\gamma},\sigma_{\gamma})
\mathcal{A}.
\end{equation}
To leading order in photon momentum the $M_1$ transition 
dominates .
We find $\Gamma = 239\times 10^{-6}\mbox{ keV}$,
which is small due to a limited phase space available for the decay. 

\subsubsection{\label{1+- to 0++}$ h_{c}(1^{+-}) \to \chi_{c0} (0^{++})
 \gamma $}
Unlike the other transitions considered so far,  the magnitude of
photon momentum in this mode is  
large {\it i.e}  of the same order of magnitude as in the other
measured magnetic dipole 
 transition $J/\psi(1^{--}) \to \eta_c(0^{-+}) \gamma$. 
The matrix element is given by 
\begin{equation}
\mathcal{M}_{N\to N'\gamma}=
\frac{e_q }{m_q}
\frac{\sqrt{3}i}{ 4\pi }
 \epsilon^{* }_{M} \cdot \left [
{\bf k} _{\gamma} \times \epsilon  ( {\bf k}_{\gamma},\sigma_{\gamma}) \right ]
\mathcal{A}.
\end{equation}
The multipole decomposition Eq.~(\ref{Dudek-multipole}) 
implies dominance of a  magnetic dipole $M_1$.  
Because of the large photon momentum, $|k_\gamma| = 100~\mbox{MeV}$  
 for this decay we find  $\Gamma =0.6\mbox{ keV}$, which is comparable with the decay  width  expected for the magnetic dipole transition $\Gamma(J/\psi \to \eta_{c} \gamma)$.

\subsubsection{\label{1+- to 0-+}$ h_{c}(1^{+-}) \to \eta_{c} (0^{-+})
 \gamma $}
This transition corresponds to the only observed transition involving
the $h_c(1^{+-})$ meson.  
The multipole decomposition Eq.~(\ref{Dudek-multipole})
implies an electric dipole $E_1$ transition. The matrix element 
can be expressed as 
\begin{equation}
\mathcal{M}_{N\to N'\gamma}
= -\frac{e_q }{m_q}
\frac{2\sqrt{3}}{4\pi}
 \epsilon^*_{M} \cdot \epsilon( {\bf k}_{\gamma}, \sigma_{\gamma})
\mathcal{D}.
\end{equation}
The experiment reports \cite{pdg-2012} $\Gamma=372~\mbox{keV}$.  The potential-quark models \cite{Eric-2005, Eichten-2002}
report a decay width in the
range $\Gamma \approx 352-498\mbox{ keV} $
and lattice \cite{Dudek-2006}
 reports $\Gamma \approx 601-663\mbox{ keV} $. 
Our model yields  $\Gamma=416\mbox{ keV}$, which
 is  consistent with these  results.

\subsubsection{\label{1++ to 1--}$ \chi_{c1}(1^{++}) \to J/\psi (1^{--})
 \gamma $}
This transition has been reported experimentally and  it was studied on the 
lattice \cite{Dudek-2006} with the later giving a central value  
 somewhat above the experimental
data albeit with a sizable error.  The results of our model seems to be in good
agreement  with experiment. 
The matrix element for this transition is given by 
\begin{equation}
\mathcal{M}_{N\to N'\gamma}
= \frac{e_q }{m_q}
\frac{\sqrt{6} i }{ 4\pi }
   \epsilon^{*}_{M} \cdot   \left [ \epsilon_{M^{\prime}}  \times 
\epsilon( {\bf k}_{\gamma} ,\sigma_{\gamma}) \right ]
  \mathcal{D}.
\end{equation}
The multipole decomposition Eq.~(\ref{Dudek-multipole})
implies the electric dipole $E_1$ and the magnetic quadrupole $M_2$
 are the two leading matrix elements. 
The  experiment reports \cite{pdg-2012}
$\Gamma =302\mbox{ keV}$. The potential
quark-models  gives  the width within the range $\Gamma \approx
215-314\mbox{ keV}$. This model yields $\Gamma=333\mbox{ keV}$.

\subsubsection{\label{0++ to 1--}$ \chi_{c0}(0^{++}) \to J/\psi (1^{--})
 \gamma $}
The multipole decomposition Eq.~(\ref{Dudek-multipole})
for this transition implies the leading transition is the dipole electric $E_1$. 
The matrix element for this transition is given by
\begin{equation}
\mathcal{M}_{N\to N'\gamma}
= \frac{e_q }{m_q}
\frac{1}{2\pi}
\epsilon( {\bf k}_{\gamma}, \sigma_{\gamma}) \cdot \epsilon_{M^{\prime}}
\mathcal{D}.
\end{equation}
The experiment reports \cite{pdg-2012}  
$\Gamma=123\mbox{ keV}$. The potential
quark-models  report  the decay width within the range $\Gamma \approx
105-152\mbox{ keV}$.  
We find  the value  $\Gamma = 265\mbox{ keV}$. 

Our results indicate approximately the same decay width
$265-363~\mbox{GeV}$   for all transitions 
that involve the charmonium multiplet $(0,1,2)^{++}$ decaying 
to the $J/\psi(1^{--})$. Experimental data  \cite{pdg-2012}, however, indicates that 
 the decay width $\Gamma(\chi_{c0} (0^{++}) \to
J/\psi \gamma)$ is approximately one third of   
$\Gamma(\chi_{c2} (2^{++})  \to J/\psi \gamma)$. 
The discrepancy is related to our simple approximation for the wave function, which ignores 
 hyperfine and  spin-orbit interactions \cite{Eichten-2002, Eric-2005}. 
 This example demonstrates that charmonium transitions can indeed be used to pin down the quark wave function.

\subsubsection{\label{1-- to 0-+}$ J/\psi(1^{--}) \to \eta_{c} (0^{-+})
 \gamma $}

This is a magnetic dipole  vector-pseudoscalar  transition between two 1S-states. 
The photon momentum for this transition is about $116~\mbox{MeV}$ and
the  transition amplitude $\hat{V}$ is calculated as is shown
in Eq.~(\ref{FV}).

The matrix element for this transition is given by
\begin{equation}
\mathcal{M}_{N\to N'\gamma}
=-\frac{e_q }{m_q}
\frac{i}{ 4\pi } \epsilon^* _{M} \cdot  \left [
{\bf k}_{\gamma} \times \epsilon( {\bf k}_{\gamma}, \sigma_{\gamma})\right ]
\mathcal{J},
\end{equation}
where $\mathcal{J}$ is defined in Appendix \ref{A:mmre}.

The experiment \cite{pdg-2012} reports $\Gamma(J/\psi \to \eta_c  \gamma)=1.5\mbox{
  keV}$ 
 for this transition. 
The potential-quark
model result is in the range $\Gamma \approx
1.9-2.9\mbox{ keV}$ and we find  $\Gamma=2.9\mbox{ keV}$. 
 
There are a few magnetic dipole transitions experimentally reported
for charmonium below the $D\bar D$-threshold ($3.73~\mbox{GeV}$), \cite{pdg-2012}. 
These are given by $J/\psi(1S) \to \eta_{c} (1S)
 \gamma $, $\psi(2S) \to \eta_{c}(2S) \gamma$ and $\psi(2S) \to
\eta_{c}(1S) \gamma$. The last two correspond to radial excitations of
S-states.  In this work, we only consider ground states for charmonium.

\subsection*{C-violating meson-to-meson transitions}

In additional to the allowed transitions, we investigated possible
charge conjugation violation matrix elements  which include:  
 $\chi_{c1} \to \eta_{c} \gamma$, $\chi_{c1} \to \chi_{c0}
\gamma$, $h_{c} \to J/\psi \gamma$, $\chi_{c2} \to \eta_{c} \gamma$, 
$\chi_{c2} \to \chi_{c0} \gamma$ and $\chi_{c2} \to \chi_{c1} \gamma$. 
 The finite matrix element for these transitions is obtained when photon is coupled to a single quark line. 
The dominant $O(k_\gamma)$ matrix elements are found for 
 $\chi_{c1} \to \chi_{c0}\gamma$ and $\chi_{c2} \to
\chi_{c1}\gamma$.

\subsubsection{$\chi_{c1}(1^{++}) \to \chi_{c0}(0^{++})\gamma$} 
The one-quark-line matrix element for this transition is given by
\begin{equation}
\mathcal{M}_{N\to N'\gamma} 
= - \frac{e_q }{m_q}
\frac{\sqrt{3} i }{ 4\pi  \sqrt{2}}  
 \epsilon^{* }_{M} \cdot  \left [ {\bf k}_{\gamma} \times  \epsilon ( {\bf k}_{\gamma},\sigma_{\gamma})
\right ] \mathcal{A} ,
\end{equation}
which according to Eq.~(\ref{Dudek-multipole}) corresponds to a
magnetic dipole $M_1$. The value obtained in this model is 
$|\hat{F}|_{CV}=0.17\mbox{ GeV}$, which corresponds to the 
magnetic dipole transition amplitude  defined by
\beqa\label{TA-CV}
|\hat{F}|_{CV}^2=\frac{1 }{2 e_q^2 }\sum_{\sigma_{\gamma},M_N, M_{N'}}
 |\mathcal{M}_{N\to N'\gamma}|^2. 
\eeqa

\subsubsection{$\chi_{c2}(2^{++}) \to \chi_{c1}(1^{++})\gamma$}
The one-quark-line matrix element for this transition is given by
\begin{equation}
\mathcal{M}_{N\to N'\gamma} =
\frac{e_q }{m_q}
\frac{3i}{ 8\pi \sqrt{2}}  \varepsilon_{ikl}  \varepsilon_{jml} 
\epsilon^{* ij}_{M} 
\epsilon^m_{M^{\prime}} 
 \left [ {\bf k}_{\gamma} \times \epsilon ( {\bf k}_{\gamma},\sigma_{\gamma}) \right ]^{k}
\mathcal{A},
\end{equation}
which according to Eq.~(\ref{Dudek-multipole}) corresponds to  $M_1$,
$E_2$ and $M_3$ transitions. 
For $|\hat{F}|_{CV}$ defined in Eq.~(\ref{TA-CV}) we obtain $|\hat{F}|_{CV}=0.10\mbox{ GeV}$.

In Table \ref{tab:MM} we summarize our findings  
and compare with the non-relativistic potential-quark model of \cite{Eric-2005}
and, when available with  the transitions amplitudes $TA=|\hat{F}|,|\hat{V}|$ from 
lattice computations \cite{Dudek-2006, Dudek-2009}. We use a single 
scale parameter for all wave functions, while in  the analysis of  lattice data 
 the scale is fitted independently for each transition. The specific values are shown in
Tab. \ref{tab:MM}. The photon momentum for each transition is given by 
$k_{\gamma}= (M^2_N -M^2_{N'})/2M_N$.

\begin{widetext}
\begin{center}
\begin{table}
\centering
\caption{Conventional $c\bar{c}$ meson transitions compared to
  NR-potential model, lattice calculations and the PDG
   values, when available. The charge
  violating transitions, described in the text  are denoted by CV. 
The input charmonium meson masses have been taken form the
  non-relativistic model of \cite{Eric-2005}. The width parameter in
the present model is fixed at  $\beta=0.5$ GeV and the ratio $R$ is defined in the text.}
\label{tab:MM}

\begin{tabular}{c cc c cc c cc c cc c cc c cc c}\hline \hline

Transition && $k_{\gamma}$ [MeV] && R  && R \cite{Eric-2005} &&
$TA$ [GeV]  &&
$(TA,~\beta)$ [GeV] \cite{Dudek-2006, Dudek-2009}   
&& $\Gamma[\mbox{keV}]$ && $\Gamma [\mbox{keV}]$\cite{pdg-2012}\\ [0.5ex] \hline

 $(\chi_{c_2} \to h_c \gamma)_{M_1} $ && 40 && $3.2\times 10^{-4}$ && - && $|\hat{F}|=0.12$
 && - && 0.1 && -\\ [0.5ex] 

 $(\chi_{c_2} \to \chi_{c1} \gamma)_{CV} $ && 45 && zero && - && $|\hat{F}|_{CV}=0.10$
 && - && zero && -\\ [0.5ex]

 $(\chi_{c_2} \to \chi_{c0} \gamma)_{CV} $ && 138 && zero && - && zero
 && - && zero && -\\ [0.5ex]

   $(\chi_{c_2} \to J/ \psi \gamma )_{E_1} $ && 429 && 1 && 1 &&
$|\hat{F}|=2.02$ && ($|\hat{F}|=1.97$, 0.55) && 363  && 380  \\ [0.5ex]

 $(\chi_{c_2} \to \eta_{c} \gamma)_{CV} $ && 530 && zero && - && zero
 && - && zero && -\\ [0.5ex]
 
  $(h_{c} \to\chi_{c_1} \gamma)_{M_1} $ && 5 && $\sim 10^{-7}$ && - &&
$|\hat{F}|= 0.01$ && - && $\sim 10^{-3}$ && -  \\ [0.5ex] 

  $(h_{c} \to\chi_{c_0} \gamma )_{M_1}$ &&100 && $1.7 \times 10^{-3}$ && -
  && $|\hat{F}|=0.13$ && - && 0.6 && - \\ [0.5ex] 

  $(h_{c} \to J/\psi \gamma )_{CV}$ && 394 && zero && -
  && zero && - && zero && - \\ [0.5ex] 

  $(h_{c} \to \eta_{c} \gamma)_{E_1} $ && 504 &&1.14 && 1.17 &&
  $|\hat{E}_1| =1.54$ && ($|\hat{E}_1| =1.85$, 0.69) && 416  && 372  \\ [0.5ex] 
 
  $(\chi_{c_1} \to \chi_{c0} \gamma)_{CV} $ && 95 && zero && - && 
$|\hat{F}|_{CV} =0.17$ && - && zero && -  \\ [0.5ex] 

  $(\chi_{c_1} \to J/\psi \gamma)_{E_1} $ && 390 && 0.92 && 0.74 && 
$|\hat{E}_1| =1.56$ && ($|\hat{E}_1|=1.88$, 0.56) && 333 && 302  \\ [0.5ex] 

  $(\chi_{c_1} \to\eta_{c} \gamma)_{CV} $ && 492 && zero &&- && 
zero &&- && zero && -  \\ [0.5ex]

 $(\chi_{c_0} \to J/\psi \gamma)_{E_1} $ && 303 && 0.73 && 0.36 && 
$|\hat{E}_1|=1.33$ && ($|\hat{E}_1|=0.83$, 0.54) && 265 && 123\\ [0.5ex] 

 $(\chi_{c_0} \to \eta_c \gamma)_{CV} $ && 408 && zero && - && 
zero && - && zero && - \\ [0.5ex]

  $( J/\psi \to \eta_c \gamma )_{M_1} $ && 116 && $7.9\times 10^{-3}$ &&
  $6.8\times 10^{-3}$ && $|\hat{V}|=1.98$/GeV && ($|\hat{V}|=1.85$/GeV,
  0.54) && 2.9 && 1.5  \\ [1ex]  
\hline  
\end{tabular}

\end{table}
\end{center}
\end{widetext}


\subsection{\label{NR:HM} Hybrid-to-meson radiative decays}

We have studied 24 possible hybrid to meson
radiative transitions, including  matrix elements for C-violating modes. 
The results are discussed below. In Table
\ref{tab:HM}, we quote the expected decay ratios for these transitions when using $m_{hyb}=4.35~\mbox{GeV}$ for the spin-averaged mass  of the lowest hybrid multiplet  $1^{--},~ (0,1,2)^{-+}$. 
 To minimize sensitivity to the wave functions we also quote the ratio
 of hybrid decay amplitudes  computed in the model, {\it cf.}
 Eq.~(\ref{FV}) to those computed using lattice simulations
 \cite{Dudek-2009}. Specifically, from lattice simulations two widths are quoted. The  $Y (1^{--}) \to \eta_c \gamma$, 
 transition from a hybrid-vector, $Y$ is a magnetic dipole with the decay width given by 
\begin{equation}{\label{hyb Y lat}}
\Gamma(Y \to \eta_c \gamma) = 
\alpha  k_\gamma ^3 
\frac{ 64 }{ 27 } \frac{ |\hat{V}|^2 }{ ( m_Y +  m_{\eta_c} )^2 } ,
\end{equation}
 where the magnetic dipole matrix element  
$\hat{V}=0.28$ yields  $\Gamma(Y \to \eta_c
\gamma)=42~\mbox{keV}$. The second transition reported in \cite{Dudek-2009}
  is $\eta_{c_1}(1^{-+}) \to J/\psi(1^{--}) \gamma$ from the exotic hybrid $\eta_{c_1}$,  which is also of a  magnetic dipole type, 
 with the decay width given by 
\begin{align}{\label{hyb eta lat}} 
\Gamma(\eta_{c_1} \to J/\psi \gamma) = 
\alpha k_\gamma \frac{ 16 }{ 27 }
\frac{ |\hat{F}|^2 }{ m_{\eta_{c1}} ^2} ,
\end{align}
 and the matrix element, $\hat{F} =
0.69~\mbox{GeV}$ gives $\Gamma(\eta_{c1} \to J/\psi
\gamma)=115~\mbox{keV}$. As it is shown below, all hybrid transition 
amplitudes in our  model depend 
 on a single factor  $|\mathcal{Z}_0|$ that is determined by the  hybrid meson wave function. 
 We will use the two magnetic dipole matrix elements, $\hat M_1$ and $\hat F$
  to normalize this factor to make  predictions for
 transitions not yet reported but calculable on our model. 
 
For the C-violating matrix elements we will use the
transition $1^{-+} \to 0^{++}\gamma$ from \cite{Dudek-2009} to normalize the relevant wave function overlap factor in our model. 
 
\subsubsection{$ Y(1^{--}) \to \eta_c (0^{-+})  \gamma $} 

This transition involves a  hybrid vector meson state denoted as $Y$. Lattice simulation of charmonium (as well as light quark mesons) predict a vector state located  between first and second resonance region {\it i.e.} above the first  radial and orbital excitation of the ground state spin-one $q\bar q$. 
Experimentally the $Y(4260)$ is a possible candidate for this
hybrid in the  charmonium spectrum  and the $Y(2175)$ is the hybrid candidate in the   $s\bar s$ sector \cite{BABAR-SS}. 
The transition between hybrid vector and ordinary pseudoscalar  $c\bar c$ meson,  is of  magnetic dipole type 
 and the matrix elements are given by
\begin{equation}
\mathcal{M}_{N \to N' \gamma} = 
 \frac{e_q   }{m_q  } 
\frac{ 3 i  }{  64 \pi^{\frac{3}{2}}}
  \epsilon^{* }_{M} \cdot   \left[ {\bf k}_{\gamma} 
\times \epsilon ( {\bf k}_{\gamma},\sigma_{\gamma}) \right]
\mathcal{Z}_{0},
\end{equation}
where $\mathcal{Z}_{0}$ involves an integral over meson wave
functions, the scalar function $K^1 (k, q)$,
the gluon absorption kernel and the Green's function $1/\Delta E$.  

The quantity $\kappa_1^2$  defined by 
\begin{equation} 
\Gamma(Y \to \eta_c \gamma)_{M_1} =\alpha { k}_{\gamma}^3    \kappa_1^{2} 
\end{equation} 
  can be compared with  the lattice result given in Eq.~(\ref{hyb Y lat}),
 \begin{equation}{\label{hyb Y us}}
 \kappa_1  =
\frac{ 1 }{ 32 \sqrt{3} \pi^{\frac{3}{2}} m_q }  
 \frac{ |\mathcal{Z}_{0}| }{m_{Y}} = \frac{ 8 }{ 3 \sqrt{3} } \frac{ |\hat{V}| }{ ( m_Y +  m_{\eta_c} ) }  ,
\end{equation}
 This gives a relation between $\mathcal{Z}_0$ and the magnetic
dipole form factor $\hat V$ and using the lattice value of  $\hat{V}=0.28$ 
we obtain  $\kappa_1^2  
= 3.47
\times 10^{-3} ~\mbox{GeV}^{-2}$
and it corresponds to  a decay width of  $\Gamma(Y \to
\eta_c \gamma)_{M_1}=40~\mbox{keV}$. 
The difference between the reported value
by lattice and this model is due to the values of the masses for the
hybrid and meson states used to calculate the photon momentum. 
The reason we take the lattice  measurement 
 to normalize  $\mathcal{Z}_0$ is because 
 of uncertainties  in its computation within the model. 
  It requires knowledge of the hybrid wave function, which in turn requires solving the three-body problem, {\it cf.}  Eq.~(\ref{sh}).  
 We leave this for future investigations and  here focus instead on symmetry relations implied by existence of the light hybrid multiplet.  
 
 Using the value for $|\mathcal{Z}_0|^2$ or equivalently  $\kappa^2_1$ estimated above, we can now 
make a prediction for the other three nonzero hybrid radiative
transitions that in our model are determined by the same wave function overlap. 
These are   $0^{-+} \to J/\psi (1^{--})
\gamma$, $\eta_{c1}(1^{-+} ) \to J/\psi (1^{--})  \gamma$ and $2^{-+}
\to J/\psi (1^{--})  \gamma$. 
The results are summarized in Table \ref{tab:HM}.

\subsubsection{$0^{-+} \to J/\psi (1^{--})  \gamma$}

This is also a dipole magnetic transition. As it is shown below, the model predicts that
any difference with respect to $Y(1^{--}) \to \eta_c (0^{-+})  \gamma $ is only due to the available phase space as determined 
 by the magnitude of the photon momentum, $k_\gamma$. 
The matrix element for this transition is given by 
\begin{equation}
\mathcal{M}_{N \to N' \gamma}  = - \frac{e_q }{m_q} 
\frac{  i  \sqrt{3} }{64 \pi^{\frac{3}{2}}}
 \epsilon_{M^{\prime}}  \cdot  \left[ { \bf k}_{\gamma}  \times
\epsilon ( {\bf k}_{\gamma},\sigma_{\gamma}) \right]
\mathcal{Z}_{0}.
\end{equation}
The normalized results for transition with respect to lattice magnetic dipole form
factors  are summarized in Table \ref{tab:HM}.

\subsubsection{$\eta_{c1}(1^{-+} ) \to J/\psi (1^{--})  \gamma$}

The multipole decomposition for this transition includes a 
  magnetic dipole and an electric quadrupole transitions but to lowest order in 
  photon momentum the magnetic dipole transition dominates. 
  The corresponding matrix element is given by
\begin{equation}
\mathcal{M}_{N \to N' \gamma}  = 
 \frac{e_q }{m_q  }  
\frac{ 3 \sqrt{2}    }{ 128 \pi^{\frac{3}{2}} }
 \left ( \epsilon^{*}_{M}  \times  \epsilon_{M^{\prime}}  \right ) \cdot \left[
\epsilon ( {\bf k}_{\gamma}, \sigma_{\gamma}) \times {\bf k}_{\gamma} \right]
\mathcal{Z}_{0},
\end{equation}
and the decay width is given by $\Gamma(\eta_{c1}  \to J/\psi  \gamma)_{M_1} =\alpha k_{\gamma} \kappa_2^2 $, where 
we have defined $\kappa_2$ as 
  \begin{equation} 
\kappa_2 \equiv  
\frac{  { k}_{\gamma}}{ 32\sqrt{3} \pi^{\frac{3}{2}} m_q }  
 \frac{ |\mathcal{Z}_{0}| }{m_{ \eta_{c1} }} =  \frac{4}{3 \sqrt{3}} \frac{|\hat{F}|}{m_{\eta_{c1}} }.
\end{equation} 
   Using the lattice value $\hat F = 0.69\mbox{ GeV}$ 
  we find 
\begin{equation} 
\kappa_2^2 =1.49 \times 10^{-2}
\end{equation} 
We can now estimate the difference in 
$|\mathcal{Z}_0|^2$ obtained using the two lattice results as
 normalizers. We find 
 \begin{equation} 
   |\mathcal{Z}_0|_{\hat{F}} \approx 2\times
|\mathcal{Z}_0|_{\hat{V}}
\end{equation} 
where the subscript indicates which lattice matrix element is used in the determination. 
  This result implies significant dependence of
$\mathcal{Z}_0$ on the process and can be interpreted  as a measure of the difference in the wave functions of the 
 $1^{--}$ and $1^{-+}$  hybrids. 
 This discrepancy is also seen in the ratio $\mathcal{R}=\Gamma(Y \to \eta_c \gamma)/\Gamma(\eta_{c_1} \to
  J/\psi \gamma)$ which is approximately $0.37$ on the lattice while our model predicts
  $\mathcal{R} \approx 1$ for the whole hybrid super multiplet. 
 In Table \ref{tab:HM}, we show the predictions for the decay widths
normalized using both  $\kappa_1$ and $\kappa_2$

\subsubsection{$ 2^{-+} \to J/\psi (1^{--})  \gamma$}
The hybrid $2^{-+}$ is the last remaining member of lightest hybrid multiplet considered here. 
Since no pseudo-tensor charmonium transition has been observed  (not even one fitting any of the
ordinary $c\bar c$ meson multiplets) , the results of this model may be relevant to  
 future experimental searches.  The matrix element corresponding to this transition is given by
\begin{equation}
\mathcal{M}_{N \to N' \gamma} = 
 \frac{e_q }{m_q  }  
\frac{  3 i  }{ 64 \pi^{\frac{3}{2}} }
 \epsilon^{* ij }_{M} \epsilon^{j}_{M^{\prime}}
 \left [{\bf k}_{\gamma}  \times \epsilon ( {\bf k}_{\gamma},\sigma_{\gamma})
\right  ]^{i} \mathcal{Z}_0 .
\end{equation}
The predictions are summarized in Table \ref{tab:HM}.

\begin{center}
\begin{table}
\caption{Expected decay widths 
for the nonzero hybrid to
meson radiative transitions. The input charmonium meson masses have
been taken form the non relativistic model in \cite{Eric-2005} and
the mass of the hybrid multiplet was set to $m_{hyb}=4.35~GeV$.}
\label{tab:HM}
\centering
\begin{tabular}{c c c c c c c}\hline \hline

Transition && $\Gamma$ for $\kappa_1$ [KeV] 
&&$\Gamma$ for $\kappa_2$ [KeV]  \\ [0.5ex] \hline 

 $(Y_1 \to \eta_{c} \gamma)_{M_1} $ && 39 && 126 \\[0.5ex]  

   $(0^{-+} \to J/ \psi \gamma )_{M_1} $&& 32 && 116 \\[0.5ex]

  $(\eta_{c_1} \to J/ \psi \gamma)_{M_1} $ && 32 && 116  \\ [0.5ex]

  $(2^{-+} \to J/ \psi \gamma )_{M_1}$ && 32 && 116 \\ [1ex] 
\hline 

\end{tabular}
\end{table}
\end{center}

\subsection*{C-violating hybrid transitions}
Two charge conjugation violating matrix elements have been reported in \cite{Dudek-2009}. Both involve the exotic hybrid and they are, 
$1^{-+} \to 0^{-+}\gamma$ and $1^{-+} \to 0^{++}\gamma$. In our model the matrix
element for the $1^{-+} \to 0^{-+} \gamma$  transition vanishes 
identically.  
The reason is that in the matrix element for the photon coupling to the quark 
{\it i.e.}  the two
terms in the curly brackets in  Eq.~(\ref{H to M})  only the spin-flip term contributes but does not bring any gluon
momentum ({\bf k}) contribution. Therefore, all the gluon momentum dependence comes from the
Coulomb interaction Eq.~(\ref{Ham-QCD-QED}) and the
hybrid wave function Eq.~(\ref{hwfs}). 
Thus, after performing the gluon angular integration Eq.~(\ref{gluon-int}), the
transition matrix element gives exactly zero from the term proportional to $\sim
\varepsilon_{ijk}  \hat{{\bf q}}_g^i \hat{{\bf q}}_g^k = 0$.

The matrix elements for the other,  non-vanishing C-violating transitions are summarized below. 
\begin{align}{\label{M-mtx 1-- to 1+-}}
\mathcal{M}_{Y \to h_c \gamma }
&= \kappa_3 e_q \frac{  3i     }{  16  (4\pi)^{ \frac{3}{2} }  }
( \epsilon^* _M  \times\epsilon_{M^\prime}  ) \cdot \epsilon ( {\bf k}_\gamma , \sigma_\gamma )
,\nn\\
\mathcal{M}_{0^{-+} \to \chi_{c1}  \gamma}  
&= \kappa_3 e_q\frac{    \sqrt{6}   }{  16  (4\pi)^{ \frac{3}{2} }  }
\epsilon ( {\bf k}_\gamma , \sigma_\gamma )
\cdot\epsilon_{M^\prime} ,\nn\\
\mathcal{M}_{ \eta_{c1} \to \chi_{c0}  \gamma}  
&= -\kappa_3 e_q\frac{    \sqrt{6}   }{  16  (4\pi)^{ \frac{3}{2} }  }
\epsilon_M^* \cdot \epsilon ( {\bf k}_\gamma , \sigma_\gamma ),  \nn\\
\mathcal{M}_{\eta_{c1} \to \chi_{c1}   \gamma}  
&=- \kappa_3 e_q\frac{   3i } {  32  (4\pi)^{ \frac{3}{2} }  }
( \epsilon^* _M  \times\epsilon_{M^\prime}  ) \cdot \epsilon ( {\bf k}_\gamma , \sigma_\gamma ) , \nn\\
\mathcal{M}_{\eta_{c1}  \to \chi_{c2} \gamma}  
&= \kappa_3 e_q\frac{   3 \sqrt{2} } {  32  (4\pi)^{ \frac{3}{2} }  }
\varepsilon_{ijk}\varepsilon_{ilm}
\epsilon_M^{*j} \epsilon^l ( {\bf k}_\gamma , \sigma_\gamma )
\epsilon^{km}_{M^\prime} ,\nn\\
\mathcal{M}_{2^{-+} \to \chi_{c1}  \gamma}  
&= \kappa_3 e_q\frac{   3 \sqrt{2} } {  32  (4\pi)^{ \frac{3}{2} }  }
\varepsilon_{ijk}\varepsilon_{lkm} \epsilon^{*il}_{M}
\epsilon^j ( {\bf k}_\gamma , \sigma_\gamma ) \epsilon_{M^\prime}^{*m} , \nn\\
\mathcal{M}_{2^{-+} \to \chi_{c2}   \gamma} 
 &= -\kappa_3 e_q 
\frac{   3i } {  16  (4\pi)^{ \frac{3}{2} }  }
\varepsilon_{ijk} \varepsilon^{*il}_{M} 
\epsilon^j ( {\bf k}_\gamma , \sigma_\gamma ) \varepsilon^{lk}_{M^\prime},
\nn\\
\end{align}
where $\kappa_3$ is defined as 
\begin{equation} 
\kappa_3  \equiv \frac{ N_c \sqrt{C_F}}{m_q}  | \mathcal{Z}_1| = \frac { 16 (4\pi)^{\frac{3}{2}} }{\sqrt{6}}  |\hat{E}_1|_{CV}.
\end{equation} 
It is observed that, as before, all matrix elements depend on a single wave function overlap factor $\mathcal{Z}_1$, 
 given in the Appendix. To determine the common factor $\kappa_3$ for all the nonzero hybrid 
C-violating transitions found in our model 
we use the transition amplitude ($|\hat{E}_1|_{CV}=0.34\mbox{GeV}$)
from lattice simulations reported for 
  $1^{-+} \to 0^{++}\gamma$. Therefore, using Eq.~(\ref{TA-CV}), 
$\kappa_3 =   98.93~\mbox{GeV}$. 
The numerical results shown in Table \ref{tab:HCV} constitute predictions of the model. 
\begin{center}
\begin{table}[h!]
\caption{C-violating expected amplitudes.}
\label{tab:HCV}
\centering
\begin{tabular}{c c c }\hline \hline

Transition &&$|\hat{E}_1|$ for $\kappa_3$ [GeV] 
 \\ [0.5ex] \hline 

 $Y \to h_{c} \gamma $ && 0.60  \\[0.5ex] 

   $0^{-+} \to \chi_{c_1} \gamma  $ && 0.34  \\[0.5ex]

  $\eta_{c_1} \to\chi_{c_0} \gamma $ && 0.34   \\ [0.5ex]

  $\eta_{c_1} \to\chi_{c_1} \gamma $ && 0.30  \\ [0.5ex]

  $\eta_{c_1} \to\chi_{c_2} \gamma $ && 0.38  \\ [0.5ex] 

  $2^{-+} \to\chi_{c_1} \gamma $ && 0.38  \\ [0.5ex] 

  $2^{-+}  \to\chi_{c_2} \gamma $ && 0.66 \\ [1ex] 
\hline 

\end{tabular}
\end{table}
\end{center}

\section{\label{SO} Summary and Outlook}

We studied radiative decays of  conventional charmonia and charmonium 
hybrids. Ordinary $c\bar{c}$-mesons with quantum numbers $J^{PC}$,
$\eta_{c}(0^{-+}), J/\psi(1^{--}), \chi_{c0}(0^{++}), \chi_{c1}(1^{++}), h_c(1^{+-}), \chi_{c2}(2^{++})$ were used as benchmark
 where we considered the minimal
coupling of the photon to the non-relativistic quarks. Simple harmonic
oscillator wave functions with fixed size parameter were used to
calculate the decay widths. We have compared our results with other models  \cite{Eichten-2002,
  Eric-2005} and found a reasonable agreement. A few new predictions for
transition amplitudes  were presented including charge violating
transitions amplitudes.

To describe hybrid decays we considered a model based on an effective QCD Hamiltonian that describes non-relativistic
quarks interacting with (relativistic) gluons 
 and is constructed from the
 QCD in the Coulomb gauge by applying Foldy-Wouthuysen 
transformation. 
We have derived all relevant matrix elements, which can be computed given a model for a hybrid meson wave function. 
   We considered decays of states from the hybrid multiplet
$1^{--},   (0; 1; 2)^{-+}$ with $1^{-+}$ being the exotic 
state. There are 24 possible radiative transitions between this multiplet and ground state charmonia. 
 The decay widths obtained
in this model  were normalized with respect to the two reported lattice
transition amplitudes for $\Gamma(Y(1^{--}) \to \eta_{c}\gamma)$ 
and  $\Gamma(\eta_{c1}(1^{-+}) \to J/\psi \gamma)$. The other two  
$\Gamma(0^{-+} \to J/\psi \gamma)$ and $\Gamma(2^{-+} \to J/\psi
\gamma)$ constitute predictions of the model. In general the model predicts 
$\mathcal{R}=\frac{\Gamma(N \to
N' \gamma)}{\Gamma(\eta_{c1}(1^{-+}) \to
J/\psi \gamma)} \approx 1$ for the whole
hybrid multiplet while lattice reports $\mathcal{R} = \frac{\Gamma(Y(1^{--}) \to
\eta_{c}\gamma)}{\Gamma(\eta_{c1}(1^{-+}) \to
J/\psi \gamma)}=0.37$.  We also investigated C-violating matrix elements 
involving hybrids   The
model predicts several of such matrix elements to be nonzero
and we used the lattice transition amplitude for $\eta_{c1}(1^{-}+) \to
0^{++}\gamma$ as normalizer to constrain out  predictions.

In absence of spin-dependent interactions, the model leads to a degenerate hybrid multiplet. 
While this prediction is not too far from lattice findings, the differences in transition matrix elements obtained from lattice simulations can be used to probe the wave functions predicted by the model.  This requires solving the hybrid meson Shr\"odingier equation. A simplified variational attempt has been made in 
\cite{Peng-Adam-2-2008} and in the future we hope to obtain a more realistic description of hybrid mesons wave functions.

\begin{acknowledgments}
This work was supported in part by 
 CONACyT under Postdoctoral supports
 No. 166115, No. 203672, the  U.S. Department of Energy under
Grant No. DE-FG0287ER40365, and Indiana University Collaborative Research Grant. P.G. and A.P.S. acknowledge support from U.S. Department of Energy contract DE-AC05-06OR23177, under which Jefferson Science Associates, LLC, manages and operates Jefferson Laboratory.
\end{acknowledgments}

\appendix

\section{\label{A:mhsowf} Meson and Hybrid spin-orbital wave
  functions}
The conventional $c\bar c$ meson spin-orbital wave functions are given by

\begin{align}{\label{mwfs}}
  &   \chi_{m_1 , m_2}^{JMPC} ( 0^{-+}  )
= \frac{1}{2\sqrt{2\pi}}  \left  [i\sigma_2 \right ]_{m_1 m_2} , \nn\\
&    \chi_{m_1 , m_2}^{JMPC} ( 1^{--} )
= \frac{1}{2\sqrt{2\pi}}   \left [\sigma
(i\sigma_2) \right]_{m_1 m_2} \cdot \epsilon_{M} , \nn\\
  &   \chi_{m_1 , m_2}^{JMPC} (0^{++})
= -\frac{1}{2\sqrt{2\pi}}  \left [\sigma
(i\sigma_2) \right ]_{m_1 m_2} \cdot \hat{ {\bf q} } , \nn\\
  &   \chi_{m_1 , m_2}^{JMPC} (1^{++})
=  -  \frac{  \sqrt{3}}{ 4 \sqrt{\pi}}  \left [\sigma
\sigma_2 \right ]_{m_1 m_2} \cdot 
\hat{ \bf{q} } \times \epsilon_M , \nn\\
 &    \chi_{m_1 , m_2}^{JMPC} (1^{+-})
= \frac{\sqrt{3}}{ 2 \sqrt{ 2\pi }}  \left   [  i\sigma_2  \right ]_{m_1 m_2}
 \hat{ {\bf q} } \cdot \epsilon_{M},  \nn\\
 &     \chi_{m_1 , m_2}^{JMPC} (2^{++})
=\frac{\sqrt{3}}{2 \sqrt{2\pi }} \sum_{ij}  \left [\sigma^{i}
( i\sigma_2 )  \right ]_{m_1 m_2}    \hat{ {\bf q} } ^j
\epsilon^{ij}_{M}. \nonumber \\
\end{align}

The hybrid spin-orbital  wave functions are given by

\begin{align} {\label{hwfs}}
 &    \chi_{m_1 , m_2 , \sigma} ^{ * JMPC} (1^{--})  \nonumber \\
 & \quad \quad  =- \frac{\sqrt{3}}{ 8\pi   }  
[i\sigma_2]_{m_2 m_1} \epsilon (-{\bf k},\sigma) \cdot \epsilon^{*}_{M}
 \left [\delta_{\sigma ,1}- \delta_{\sigma, -1} \right ] , \nn\\
 &  \chi_{m_1 , m_2 , \sigma} ^{* JMPC} (0^{-+})  \nonumber \\
 &\quad \quad  = \frac{1}{8\pi}  \left    [( i \sigma_2) \sigma \right ]_{m_2 m_1} \cdot   \epsilon(-{\bf k},\sigma)
\left  [\delta_{\sigma, 1}- \delta_{\sigma, -1} \right ] , \nn\\
  &    \chi_{m_1 , m_2 , \sigma} ^{* JMPC} 
(1^{-+})  \nonumber \\
&\quad \quad = \frac{\sqrt{3}}{ 8\pi \sqrt{2}}  
 \left  [\sigma_2 \sigma \right ]_{m_2 m_1} 
  \cdot \epsilon(-{\bf k},\sigma ) \times \epsilon^{*}_{M}\left
 [\delta_{\sigma, 1}- \delta_{\sigma ,-1} \right]  ,\nn\\
 &   \chi_{m_1 ,m_2 \sigma} ^{*JMPC}(2^{-+} ) 
 \nonumber \\
 & \quad = -\frac{\sqrt{3}}{8\pi }  
\sum_{ij}   \left  [ ( i \sigma_2 ) \sigma^{j}
 \right ]_{m_2 m_1}  \epsilon^{i}(-{\bf k}, \sigma) 
 \epsilon^{* ij}_{M} \left
[\delta_{\sigma ,1} - \delta_{\sigma,-1} \right ]  .\nn\\
\end{align}

\section{{\label{A:mmre}}Meson to meson relevant expressions}
The radiative transitions between two conventional mesons produce the
following set integrations 
\begin{align}
&\int \frac{d {\bf q}}{(2\pi)^3}   \Psi_{c\bar{c}}^{N,\alpha}(q)
\Psi_{c\bar{c}}^{N^{\prime},\alpha^{\prime}} (q) 
= \mathcal{J}, \nn\\
&\int \frac{d {\bf q}}{(2\pi)^3}   \Psi_{c\bar{c}}^{N,\alpha}(q)
\Psi_{c\bar{c}}^{N^{\prime},\alpha^{\prime}} (|{\bf q}- \frac{{\bf
  k}_{\gamma}}{2}|)  \hat{{\bf q}}^i \hat{{\bf q}}^j  
= \mathcal{A} \delta_{ij} 
+ \mathcal{B} \hat{{\bf k}}^i_{\gamma} \hat{{\bf k}}^j_{\gamma} , \nn\\
&\int \frac{d {\bf q}}{(2\pi)^3}   \Psi_{c\bar{c}}^{N,\alpha}(q)
\Psi_{c\bar{c}}^{N^{\prime},\alpha^{\prime}} (|{\bf q}- \frac{{\bf
  k}_{\gamma}}{2}|) { q} \hat{{\bf q}}^i \hat{{\bf q}}^j  
= \mathcal{D} \delta_{ij} 
+ \mathcal{G} \hat{{\bf k}}^i_{\gamma} \hat{ {\bf k}}^j_{\gamma} , \nn\\
\end{align}
and
\begin{align}
\mathcal{A}&=\int \frac{d {\bf q}}{(2\pi)^3}   \Psi_{c\bar{c}}^{N,\alpha}(q)
\Psi_{c\bar{c}}^{N^{\prime},\alpha^{\prime}} (|{\bf q}-\frac{{\bf
  k}_{\gamma}}{2}|) 
\left(\frac{1-\hat{y}^2}{2}\right)\nn\\
\mathcal{B}&=\int \frac{d  {\bf q}}{(2\pi)^3}   \Psi_{c\bar{c}}^{N,\alpha}(q)
\Psi_{c\bar{c}}^{N^{\prime},\alpha^{\prime}} (|{\bf q}-\frac{{\bf
  k}_{\gamma}}{2}|) 
\left(\frac{3\hat{y}^2 -1}{2}\right),\nn\\
\mathcal{D}&=\int \frac{d  {\bf q}}{(2\pi)^3}   \Psi_{c\bar{c}}^{N,\alpha}(q)
\Psi_{c\bar{c}}^{N^{\prime},\alpha^{\prime}} (|{\bf q}-\frac{{\bf
  k}_{\gamma}}{2}|) { q}
\left(\frac{1-\hat{y}^2}{2}\right), \nn\\
\mathcal{G}&=\int \frac{d  {\bf q}}{(2\pi)^3}   \Psi_{c\bar{c}}^{N,\alpha}(q)
\Psi_{c\bar{c}}^{N^{\prime},\alpha^{\prime}} (|{\bf q}-\frac{{\bf
  k}_{\gamma}}{2}|)  { q}
\left(\frac{3\hat{y}^2 -1}{2}\right), \nn\\
\end{align}
 where $\hat{y} = \hat{{\bf q}} \cdot \hat{{\bf k}}_{\gamma}$.

\section{{\label{A:hmre}}Hybrid to meson relevant expressions}

In the hybrid to meson radiative transition, the integrals over the direction of gluon momentum produce the
following set of relations
\begin{align}\label{gluon-int}
&\int d \hat{{\bf k}} K^{(1)}(|\frac{{\bf k}}{2} + {\bf q}_g|, 
|\frac{{\bf k}}{2} - {\bf q}_g |) 
\hat{ {\bf k}}^i  \hat{ {\bf k}}^j \nn\\
& \quad \quad  \quad \quad \quad = \mathcal{A}(k, q_g)\delta_{ij} 
+\mathcal{B}(k, q_g)\hat{{\bf q}}^{i}_g \hat{{\bf q}}^{j}_g , 
\end{align}
with ${\bf q}_g =( {\bf
  q}^{\prime} - {\bf q} + \frac{ {\bf k}_{\gamma} }{2} ) $
and $x=\hat{{\bf q}}_g \cdot \hat{{\bf k}}$
\begin{align}
\mathcal{A}(k,q_g)&=\int d \hat{{\bf k}}  K^{(1)}(|\frac{{\bf k}}{2} + {\bf q}_g|, 
|\frac{{\bf k}}{2} - {\bf q}_g |) \frac{1-x^2}{2}, \nn\\
\mathcal{B}(k,q_g)&=\int d \hat{{\bf k}}    K^{(1)}(|\frac{{\bf k}}{2} + {\bf q}_g|, 
|\frac{{\bf k}}{2} - {\bf q}_g |) \frac{3x^2 -1}{2} . \nn\\
\end{align}
To leading order in photon momentum ($q_g \to |{\bf q}^{\prime} -{\bf
  q}|$)  we use the following notation
\begin{align}\label{Z0}
\mathcal{Z}_{0}
&= g^2\int \frac{ k^2 d k}{(2\pi)^3}  \frac{ d {\bf q}}{(2\pi)^3}
\frac{   d  {\bf q}^{\prime}}{(2\pi)^3} 
\Psi_{c\bar{c}g}^{N,\alpha}(k,q)
\Psi_{c\bar{c}}^{N^{\prime},\alpha^{\prime}} (q^{\prime}) \nn\\
&\times \frac{ k }{ \sqrt{ \omega_k} (\Delta E) }
\mathcal{A}(k,|{\bf q}^{\prime}-{\bf q}|).
\end{align}
 The dependence on the QCD coupling $g^2$ is a consequence of the presence of the gluon in the hybrid meson wave function 
as discussed in Section~\ref{Ham}.

\subsection*{{\label{A:hmC-vr}}Hybrid to meson C-violating relations}
The C-violating transitions are given by terms proportional
to $\mathcal{A}(  { k}  , | {\bf
  q}^{\prime} - {\bf q} | )  ( q^{\prime } - q)^i \hat{ {\bf q}}^{\prime j }
$, thus,   the expressions in the C-violating hybrid to meson transitions can be simplified to
\begin{align}
\mathcal{Z}_{1}  
&= \frac{g^2}{3}\int \frac{ k^2 d k}{(2\pi)^3}  \frac{ d  {\bf q}}{(2\pi)^3}
\frac{ d  {\bf q}^{\prime}}{(2\pi)^3} 
\Psi_{c\bar{c}g}^{N,\alpha}(k,q)
\Psi_{c\bar{c}}^{N^{\prime},\alpha^{\prime}} (q^{\prime}) \nn\\
&\times\frac{k}{ \sqrt{ \omega_k } (\Delta E)}\mathcal{A}(k,|{\bf q}^{\prime}-{\bf q}|) 
\left(  { q}^{\prime} - q  \hat{z}  \right) .
\end{align}

\section{\label{A:mdwd} Multipole decomposition and 
width decay}

We need to determine the type of transition through the multipole
decomposition. The simplest way is to consider that the photon moves
in the $-\hat{{\bf z}}$ direction as in \cite{Dudek-2006}, so that,
the multipole decomposition is given by
\begin{align}{\label{Dudek-multipole}}
\mathcal{M}(\lambda_{\gamma}=\pm)
&=\sum_{l}\sqrt{ \frac{2l+1}{2J+1} }
\langle l\mp1, J^{\prime}\lambda\pm1 | J\lambda \rangle \nn\\
&\times[E_l \frac{1}{2}(1+(-1)^{l} \delta P)\mp M_l \frac{1}{2}(1-(-1)^{l}
\delta P) ],
\nn\\
\end{align}
 where the transition can be represented
as $(J\lambda)\rightarrow
(J^{\prime}\lambda^{\prime})+(\gamma\lambda_{\gamma})$,
and $\delta P$ is the product of the initial and final meson
parities.


\begin{thebibliography}{99}


\bibitem{Horn} D. Horn and J. Mandula, Phys. Rev. D, \bd{17},
  898, (1978).

\bibitem{Isgur} N. Isgur and J. E. Paton, Phys. Rev. D {\bf 31}, 2910 (1985).

\bibitem{Simonov} Y. A. Simonov, Nucl. Phys. B592, 350 (2001).


\bibitem{Adam-Eric-1996} A. Szczepaniak, E. S. Swanson, C. R. Ji, and S. R.
Cotanch, Phys. Rev. Lett. 76, 2011 (1996).



\bibitem{Buisseret}  F. Buisseret and C. Semay, Phys. Rev. D 74, 114018
(2006).

\bibitem{Brau} F. Brau and C. Semay, Phys. Rev. D 70, 014017 (2004).

\bibitem{glueball-1} C. J. Morningstar and M. J. Peardon, 
Phys.\ Rev.\ D {\bf 60}, 034509 (1999).

\bibitem{glueball-2} M. Foster and C. Michael, 
Phys.\ Rev.\ D {\bf 59}, 094509 (1999).

\bibitem{glueball-3} G.S. Bali and A. Pineda,
Phys. Rev. D {\bf 69}, 0944001, (2004)


\bibitem{Cornell} E. Eichten and F. Feinberg, 
Phys. Rev. D {\bf 23},2724 (1981).










\bibitem{Lee}  T. D. Lee, Particle Physics And Introduction To Field Theory
~Harwood Academic, New York, 1981.

\bibitem{Llanes} F. J. Llanes-Estrada and S. R. Cotanch, Phys. Lett. B 504,
15 (2001).

\bibitem{General} I. J. General, S. R. Cotanch, and F. J. Llanes-Estrada, Eur.
Phys. J. C 51, 347 (2007).

\bibitem{Peng-Adam-1-2008} P. Guo, Adam P. Szczepaniak, G.
  Galata, A. Vassallo and E. Santopinto, Phys.\ Rev.\ D {\bf 77}, 056005 (2008).

\bibitem{Peng-Adam-2-2008} P. Guo, Adam P. Szczepaniak, G.
  Galata, A. Vassallo and E. Santopinto, Phys.\ Rev.\ D {\bf 78}, 056003 (2008).







\bibitem{Feuchter1} C. Feuchter and H. Reinhardt, Phys.\ Rev.\ D {\bf 70}, 105021
(2004).

\bibitem{Feuchter2} H. Reinhardt and C. Feuchter, Phys.\ Rev.\ D {\bf 71}, 105002
(2005).



\bibitem{Zwanziger} D.  Zwanziger, Phys.\ Rev.\ Lett.\ {\bf 90},
  102001 (2003).



\bibitem{Greensite} J. Greensite and S. Olejnik, Phys. Rev. D {\bf
    67}, 094503 (2003).




\bibitem{glueball-rev1}   V. Crede and  C. A. Meyer,
Prog. Part. Nucl. Phys. {\bf 63}, 74, (2009).

\bibitem{glueball-rev2} V. Mathieu, N. Kochelev and V. Vento,
Int. J. Mod. Phys. E {\bf 18}, 1, (2009).

\bibitem{Eric-2006}  E. S. Swanson,
  Phys. Rep.   \bd{429}, 243, (2006).




\bibitem{Eichten-1978} E. Eichten, K. Gottfried, T. Kinoshita,
  K. D. Lane  and T. M. Yan, 
Phys. Rev. D {\bf 17}, 3090, (1978).

\bibitem{Eichten-1980}  E. Eichten, K. Gottfried, T. Kinoshita,
  K. D. Lane  and T. M. Yan, Phys. Rev. D {\bf 21}, 203, (1980).


\bibitem{Eichten-2002} E. J. Eichten, K. Lane and C. Quigg, 
Phys. Rev. Lett. \bd{89}, 162002, (2002).

\bibitem{Eric-2005}  T. Barnes, S. Godfrey and E. S. Swanson,
  Phys. Rev. D  \bd{72}, 054026, (2005).



\bibitem{Dudek-2006} J. J. Dudek, R. G. Edwards, and D.
  G. Richards, Phys. Rev. D \bd{73}, 074507, (2006).

 
\bibitem{Dudek-2008}  J. J. Dudek, R. G. Edwards, N.
  Mathur, and D. G. Richards, Phys. Rev. D \bd{77}, 034501, (2008)





\bibitem{Dudek-2009} J. J. Dudek, R. G. Edwards, and
  C. E. Thomas, Phys. Rev. D \bd{79}, 094504, (2009).


\bibitem{Dudek-2011}  J. J. Dudek, Phys. Rev. D \bd{84}, 074023, (2011).









\bibitem{Foldy-1978} F. L. Feinberg,  Phys.\ Rev.\ D {\bf 17}, 2659 (1978).

\bibitem{Adam-Eric-2001} A. P. Szczepaniak and E. S. Swanson, Phys.\ Rev.\ D {\bf 65}, 025012 (2001).



\bibitem{Adam-2004} A. P. Szczepaniak, Phys.\ Rev.\ D {\bf 69}, 074031 (2004).


\bibitem{Hugo-2004} C. Feuchter and H. Reinhardt, Phys. Rev. D {\bf 70}, 105021 
(2004).

\bibitem{Hugo-2005} H. Reinhardt and C. Feuchter, Phys. Rev. D {\bf 71}, 105002
(2005).

\bibitem{Hugo-2006} W. Schleifenbaum, M. Leder, and H. Reinhardt, Phys.
Rev. D {\bf 73}, 125019 (2006).

\bibitem{Hugo-2007} D. Epple, H. Reinhardt, and W. Schleifenbaum, Phys. Rev.
D {\bf 75}, 045011 (2007).

\bibitem{Hugo-Adam-2008} D. Epple, H. Reinhardt, W. Schleifenbaum, and A. P.
Szczepaniak, Phys. Rev. D 77, 085007 (2008).



\bibitem{Krupinski 1} A. P. Szczepaniak and P. Krupinski, Phys. Rev. D
  {\bf 73},
116002 (2006).

\bibitem{Krupinski 2} A. P. Szczepaniak and P. Krupinski, Phys. Rev. D
  {\bf 73},
034022 (2006).




\bibitem{Morningstar-1998} K.J. Juge, J. Kuti, and C.J. Morningstar,  Nucl. \ Phys.\ B (Proc. Suppl) {\bf 63}, 326 (1998).



\bibitem{Gu-1999} Y. F. Gu and S.F. Tuan, arXiv:9910423 [hep-ph]



\bibitem{cern-col}  N. Brambilla et al., arXiv:hep-ph/0412158v2   


\bibitem{pdg-2012} J. Beringer et al. (Particle Data Group),
  Phys. Rev. D {\bf 86}, 010001 (2012).



\bibitem{E760} E760 Collaboration, T.A. Armstrong et al.,
  Nucl. Phys. B{\bf 373}, 35 (1992); Phys. Rev. Lett.
{\bf 68}, 1468 (1992).

\bibitem{BES} BES Collaboration, J. Z. Bai et al.,
Phys. Rev. Lett. {\bf 81}, 3091 (1998); Phys. Rev. D {\bf 60},
072001 (1999);.

\bibitem{CLEO} N. E. Adam et al., Phys. Rev. Lett. {\bf 94},  232002
  (2005).



\bibitem{BABAR-SS}  B. Aubert et al. (BABAR), Phys. Rev. D {\bf 74}, 091103
(2006).

\end{thebibliography}
\end{document}